\documentstyle[12pt]{article}
\advance\textheight by \topskip
\newcommand{\sla}{\kern -5.4pt /}
\newcommand{\Dir}{\kern -6.4pt\Big{/}}
\newcommand{\Dirin}{\kern -10.4pt\Big{/}\kern 4.4pt}
\newcommand{\DDir}{\kern -7.6pt\Big{/}}
\newcommand{\DGir}{\kern -6.0pt\Big{/}}

\newcommand{\ra}{\rightarrow}
\newcommand{\be}{\begin{equation}}
\newcommand{\ee}{\end{equation}}
\newcommand{\bea}{\begin{eqnarray}}
\newcommand{\eea}{\end{eqnarray}}
\newcommand{\beanon}{\begin{eqnarray*}}
\newcommand{\eeanon}{\end{eqnarray*}}
\newcommand{\ba}{\begin{array}}
\newcommand{\ea}{\end{array}}
\newcommand{\bi}{\begin{itemize}}
\newcommand{\ei}{\end{itemize}}
\newcommand{\ben}{\begin{enumerate}}
\newcommand{\een}{\end{enumerate}}
\newcommand{\bc}{\begin{center}}
\newcommand{\ec}{\end{center}}

\newcommand{\ar}{\rightarrow}

\newcommand{\NP}[1]{{\it Nucl.\ Phys.\ }{\bf #1}}
\newcommand{\PL}[1]{{\it Phys.\ Lett.\ }{\bf #1}}
\newcommand{\ZP}[1]{{\it Z.\ Phys.\ }{\bf #1}}

\newcommand{\PR}[1]{{\it Phys.\ Rev.\ }{\bf #1}}

\begin{document}
\tolerance=100000
\input feynman
\thispagestyle{empty}
\setcounter{page}{0}

\begin{flushright}
{\large DFTT 32/96}\\ 
{\rm July 1996\hspace*{.5 truecm}}\\ 
hep-ph/9607288
\end{flushright}

\vspace*{\fill}

\bc     
{\Large \bf $b \bar b W^+ W^-$ production at hadron colliders. Top signal and
irreducible backgrounds 
\footnote{ Work supported in part by Ministero 
dell' Universit\`a e della Ricerca Scientifica.\\[2 mm]
e-mail: ballestrero@to.infn.it,maina@to.infn.it,pizzio@to.infn.it}}\\[2.cm]
{\large Alessandro Ballestrero, Ezio Maina, Marco Pizzio}\\[.3 cm]
{\it Dipartimento di Fisica Teorica, Universit\`a di Torino, Italy}\\
{\it and INFN, Sezione di Torino, Italy}\\
{\it v. Giuria 1, 10125 Torino, Italy.}\\
\ec

\vspace*{\fill}

\begin{abstract}
{\normalsize
\noindent
We compute complete tree level matrix elements for $gg , q \bar q \rightarrow
b \bar b W^+W^-$. We analyze the irreducible backgrounds to top signal 
at the Tevatron and at the LHC. Their contribution to the total cross section
is about $5 \%$ at the LHC, due to single resonant channels.
Several distributions
with contributions from signal and backgrounds are presented.}
\end{abstract}

\vspace*{\fill}
\newpage
\subsection*{Introduction}
With the discovery of the top quark at the Tevatron \cite{topdisc} by the CDF
and D0 collaborations, all fermions
belonging to the three generations which, according to the LEP measurement of
the invisible width of the $Z$, possess a light neutrino have been
observed. 
\par
Using 67 $pb^{-1}$ of data CDF measured the top mass to be
$M_t = 176 \ \pm 8 ({\rm stat}) \ \pm 10 ({\rm syst}) \  {\rm GeV}$
and the production cross section as $\sigma_{tt} = 6.8^{+3.6}_{-2.4}\ pb$.
The larger sample of top events which has been subsequently collected 
will soon allow a better determination of the parameters and properties of the
newly discovered particle.
Already at this year winter conferences new preliminary results have been
presented, based on about 110 $pb^{-1}$ of data \cite{Moriond}. The most
precise measurement of the top mass, extracted from the lepton plus three or
more jets sample, is 
$M_t = 175.6 \ \pm 5.7 ({\rm stat}) \ \pm 7.1 ({\rm syst}) \  {\rm GeV}$.
The corresponding result for the production cross section is
$\sigma_{tt} = 7.5^{+1.9}_{-1.6}\ pb$.
\par
The mass of the top is a fundamental parameter of the Standard Model and plays
a crucial role in radiative corrections to electroweak observables. 
In fact bounds on the top mass can be obtained from a comparison of high
precision measurements at LEP and SLC,
together with information from neutrino scattering, with the best theoretical
predictions. Within the Standard Model the top mass must lie in the range
$M_t = 178  \pm 8^{+17}_{-20}$ GeV 
\cite{Bruxelles}. The agreement between this indirect determination
of $M_t$ with the value which is directly measured at Fermilab
is a remarkable success of the model. 
Any improvement in the precision of the top mass measurement,
particularly when the mass of the $W$ boson will be
measured to a precision of about 50 MeV at LEP 2, will further test
the electroweak theory, significantly reducing the allowed range for the mass
of the Higgs boson.
\par 
It is obvious that the discovery of the top is only the first step.
All details of the production process and decay will 
have to be thoroughly examined.
The measurement of the production cross section and of the distributions
of different kinematic variables, like the $p_T$ of the top, the angular
distribution of the decay products or the characteristics of additional
gluon jets, will challenge our understanding of perturbative QCD.
The determination of the top decay channels will test whether the new heavy
quark behaves as predicted by the Standard Model and might provide a window on
new physics. 
These studies will be continued at Fermilab after the construction of the Main
Injector and will be further refined at the Large Hadron Collider (LHC),
where the cross section is much larger than at Tevatron energy and at 
the Next Linear Collider (NLC), in the clean environment of an $e^+e^-$ machine.
\par
In order to achieve a complete understanding of top production, a theoretical
effort matching the advances on the experimental side is required.
The production cross section has long been computed at 
next to Leading Order (NLO)
in QCD \cite{NLO}. More recently the contribution of soft gluons have been
resummed to all order \cite{reNLO} using different techniques. 
In ref.~\cite{rePt} the resummation procedure has been applied to the
inclusive transverse momentum and rapidity distributions of the top.
Several kinematic distributions have been computed at NLO in 
ref.~\cite{frix}.
All these studies have treated the top quark as a stable particle,
separating the production of a $\bar t t$ pair from the independent decay
of the two heavy quarks, in the spirit of the so--called
narrow width approximation (NWA).
\par
The overall uncertainty in the calculation of the total cross section due to
different choices of parton distribution functions and of renormalization and
factorization scales is of the order of 20\%. The difference between pure NLO
calculations and calculations which also resum soft gluon effects is also of
order 20\% at the Tevatron and as large as a factor of two at the LHC. These
uncertainties however affect in a similar way all the contributions to the
total cross section, including irreducible backgrounds whose relevance remains
to be assessed.
\par
In this letter we examine the production of the $b \bar b W^+W^-$ final state
at the Tevatron and at the LHC, taking into account the full set of
tree level diagrams. 
Our approach allows us to study the effects of the finite width of the top
and of the irreducible background to $\bar t t$ production, including their
interference.
Correlations between the decays of the two $t$--quarks are automatically
included.
\par
A few representative diagrams are shown in figs.~1 and 2.
In addition to the standard mechanism which is ${\cal O}(\alpha^2 \alpha_s^2)$
we have studied the contribution of the ${\cal O}(\alpha^4 )$ diagrams in
which an initial pair of light quarks annihilates to a photon or a $Z$ boson.
Many of the diagrams we are studying (3,4,5 in fig.~1; 2 in fig.~2), while
contributing to the irreducible background to $t\bar t$ production,
effectively represent the production of a {\it single} top and must be taken
into account in any attempt to study this rarer top production mechanism which
is of interest in his own right.
The simpler process $q \bar q \rightarrow W^* \rightarrow tb$, for instance,
has been discussed in ref~\cite{stwill} as a promising candidate for
a precise measurement of the $V_{tb}$ element of the CKM mixing matrix. 
In ref.~\cite{KS} the subset of diagrams (1,2 plus crossed in fig.~1
and 1 in fig.~2) which describe the production of a $\bar t t$ pair 
$ \bar q q, \ gg  \rightarrow \bar t t$ followed by the two decays
$t\rightarrow W^+ b$, $\bar t\rightarrow W^- \bar b$ has been studied using
helicity amplitude methods. Strictly speaking, since the three diagrams 
discussed
in ref.~\cite{KS} are not by themselves gauge invariant, the results
obtained in this fashion are doubtful.
It can however be argued that using 
an appropriate gauge, and under experimental conditions optimized
for the observation of on--shell top pairs, the error 
will be of order $\Gamma_t/M_t$.
Here we extend the helicity amplitude approach to the full gauge
invariant set of diagrams which are required to describe $b \bar b W^+W^-$
production.
\par
Within the present range of
experimentally allowed top mass values \cite{topdisc}, the top lifetime is of 
the order of $10^{-23}$ $sec$ and the top decays  to 
$b W^+$ before hadronizing.  As a consequence, there might be 
significant finite width and irreducible background effects at the percent
level. Indeed we have previously found corrections of this order of magnitude
in $e^+e^-\rightarrow b\bar bW^+W^-$ at the NLC \cite{noiee}. It is therefore
important to evaluate these effects 
and, when relevant, fully include them in the theoretical analysis.
The determination of the top mass and of the production
cross section are clearly influenced by irreducible
backgrounds and one cannot rely on the NWA without checking the size of the
corrections to this approximation. Furthermore, signal and background usually
have different distributions in phase space and the spectra of some interesting
kinematical observables may be more sensitive to these corrections than
the total cross section. Alternatively, these differences may be useful for a
more effective separation of the top signal from non resonant backgrounds.
\par
We have made no approximation concerning the mass of the $b$, taking it into
full account both in  phase space and into the computation of the matrix
elements. The $b$ mass also regulates the collinear singularities that would be
present in some of the non resonant diagrams, 
for instance diagram 3, 4, 6 and 7
in fig.~1, if we had treated the $b$ quark as massless.

\subsection*{Calculation}
The matrix elements for $gg \ar W^+W^-\bar b b$ and for
$q \bar q \ar W^+W^-\bar b b$ have been computed using helicity amplitude 
methods. For the first process the method  of ref.~\cite{hz}  
has been used. The amplitude for the second reaction has been computed with 
the method described in ref.~\cite{method}. The expression of the corresponding
Feynman diagrams has been generated with the help of PHACT 
(Program for Helicity Amplitude Calculations with Tau matrices)
\cite{phact}. 
We have checked our results for gauge and BRST invariance \cite{BRST}.
\par
The full set of
diagrams has been divided into four subsets, which were named after the 
corresponding structure of resonant enhancements as double resonant,
$t$ single resonant, $\bar t$ single resonant, and non resonant.
\par 
Some representative
diagrams for $gg \ar W^+W^-\bar b b$ are shown in fig.~1. The full set
includes 39 ${\cal O}(\alpha^2 \alpha_s^2)$ diagrams 
including 8 diagrams with a virtual Higgs. These latter are suppressed by the
small coupling of the Higgs boson to the $b$ and can be neglected.
Diagrams $1$ and $2$ are double resonant.
They correspond in narrow width approximation to the production and decay
of a $t \bar t$ pair. Integrating over the whole phase space the amplitude
squared obtained from just these two diagrams, one reproduces in NWA the cross
section for $gg \ar \bar t t$. The branching ratio of $t$ ($\bar t$) to 
$W^+ b$ ($W^-\bar b$) is in fact one, as we  take off-diagonal CKM matrix 
elements to be zero. 
\par
Diagrams $3,4$ and $5$ are examples of diagrams which are single resonant
in the $t$ channel. They correspond
in narrow width approximation to the production of a $t$--quark together
with a $\bar b W^-$ pair followed by the
decay $t \ar b W^+$. All diagrams in the same class can be 
obtained permuting gluon lines and connecting the negatively charged 
$W$ in all possible positions, while leaving the $W^+$ next to the outgoing
fermion. 
The single resonant diagrams corresponding to the production
of a $\bar t$ particle can be easily obtained in a similar way.
The remaining diagrams in fig.~1, numbered $6,7$ and $8$, are examples of non
resonant diagrams.
\par
Selected diagrams describing $q \bar q \ar W^+W^-\bar b b$ are shown in fig.~2.
The full set includes 16 ${\cal O}(\alpha^2 \alpha_s^2)$ diagrams
including 2 diagrams with a virtual Higgs and 62
${\cal O}(\alpha^4 )$ diagrams, of which 11 include a Higgs. 
In the former case the diagrams with a Higgs propagator are suppressed by the
$Hb\bar b$ coupling. In the latter case the Higgs can be connected with a large
coupling to the intermediate $Z$. However the contribution of these diagrams
can be relevant only if the Higgs mass is larger than $2M_W$ and the Higgs 
propagator can go on mass--shell. To simplify our argument,
we have neglected all diagrams which include a Higgs particle, limiting our
discussion to the diagrams which survive in the limit of an infinitely heavy
Higgs. In case a Higgs is discovered, the complete set of diagrams would have to
be considered. For a similar discussion at the NLC see \cite{noiee, noieeH}.
In analogy with the $gg \ar W^+W^-\bar b b$ case, the full set of diagrams
can be subdivided in a double resonant ($1$ of fig.~2) subset, two single
resonant subsets corresponding to $t$ production ($2$ of fig.~2) and $\bar t$ 
production and a non resonant subset ($3,4,5,6$).
\par
We have found it convenient to separate the different contributions to the cross
section and to integrate them separately, adapting the choice of phase space
variables to the peaking structure which correspond to the different
sets of intermediate particles which can go on mass shell. 
We have identified a double resonant contribution generated by
the modulus square of the corresponding diagrams and two single resonant
contributions which include the modulus square of the single resonant diagrams
plus the interference between single and double resonant diagrams. 
All remaining terms, the modulus square of non resonant diagrams plus
the interference between single resonant and non--resonant diagrams,
give the last, non--resonant, contribution.
For both reactions the four terms have been numerically integrated
using Vegas \cite{vegas} and the resulting distributions have been summed in
the end.
\par
The relative size of the different contributions is obviously dependent on the
gauge and only the sum of all terms is gauge independent. 
In a given gauge however one can meaningfully discuss the numerical significance
of the various expressions. In particular, in the physical, purely transverse
gauge we have employed, there is a clear hierarchy between the four
contributions in which we have divided the full cross section. The double
resonant contribution is always much larger than the single resonant ones, which
in turn dominate the non resonant term.
\par
We have  verified numerically that in NWA the total cross sections calculated
from the double resonant subsets reproduce the results obtained for
$gg \ar \bar t t$ and $q \bar q \ar \bar t t$ for on shell top--quarks. 
\par
We have used the MRS(D$_-$) \cite{mrsd}
set of parton distribution functions throughout this
paper. For $\alpha_s$ we have used the one--loop expression with 
$Q^2 = M_t^2$. A different choice of parton distribution functions
or of the scale at which $\alpha_s$ is evaluated would slightly alter our
results for the various cross section. However we are here mainly interested
in the {\it relative} weight of the different contributions and their ratio is
essentially insensitive to these choices.
\par

\subsection*{Results}
Our results for the total cross section are presented in table~1 and 
table~2. For the Tevatron we have studied only the energy available at present,
$\sqrt{s} = 1.8$ TeV, neglecting possible improvements that have been recently
discussed in the literature. The LHC will probably begin its operation with a
center of mass energy of 10 TeV which will be raised after a few years of
running to $\sqrt{s} = 14$ TeV. Both
energies have thus been examined. For the top mass we have chosen three
values which bracket the allowed range of variation, namely $M_t = 150$ GeV,
$M_t = 175$ GeV and $M_t = 200$ GeV.
In table~1 we report the value of the total cross section, separating the
contribution of the $gg$ and $q\bar q$ channels. In the first case we further
differentiate between the double resonant contribution, corresponding to the
production of two nearly on--shell top quarks, and the background which is
dominated by events in which only one top is produced close to its 
mass shell. In the $q\bar q$ channel we distinguish the standard mechanism
in which the initial pair of light quarks annihilates to a gluon from 
the ${\cal O}(\alpha^4 )$ contribution in
which the $q \bar q$ pair annihilates to a photon or a $Z$ boson. In both
$q \bar q$--initiated processes all non double resonant contributions are
negligible.
The rightmost column gives the sum of all partial results.
All contributions not mentioned in table~1 are negligible, well below the
percent level.
\par
Table~1 shows, as well known, that
$q\bar q$ annihilation is the dominant source of $b \bar b
W^+W^-$ events at the Tevatron, while the $gg$ fusion channel contributes about
20\% for $M_t = 150$ GeV and only about 7\% for $M_t = 200$ GeV, due to the
softer gluon spectrum at these energies.
At the LHC this pattern is reversed, the gluon fusion mechanism provides
the bulk 
of all events while the $q\bar q$ channel contributes between 10 and 20\%
depending on the top mass. We notice that the ${\cal O}(\alpha^4 )$
$q\bar q$ annihilation cross section is about 2.5\% of the corresponding
${\cal O}(\alpha_s^2\alpha^2 )$ results,
independently of the top mass and of the
collider energy. The gluon fusion background gives an important contribution 
to the cross section for $ gg \rightarrow b \bar b W^+W^-$ at the
Tevatron, between 6\% for $M_t = 150$ GeV and 18\% for $M_t = 200$ GeV.
However it only contributes about 1\% to the total cross section for $t\bar t$
production.
At the LHC the $gg$ background is smaller compared to the $gg$ signal than at
the Tevatron, but its contribution to the total $t\bar t$ cross section is
larger.
This can be appreciated in 
table~2 where we present the contribution in percent to the total cross section
at the LHC of the three main subprocesses, the double resonant channel
$gg \rightarrow t \bar t \rightarrow b \bar b W^+W^-$, the single resonant
channel 
$gg \rightarrow t \bar b W^- , \bar t b W^+ \rightarrow b \bar b W^+W^-$
and the double resonant $q \bar q$ contribution
$q \bar q \rightarrow t \bar t \rightarrow b \bar b W^+W^-$, for the two
energy values and the three values of the top mass which we have studied.
The gluon fusion background contributes between 4\% and 6\% of the total $gg$
cross section, the largest contribution being for the heaviest top mass.
Comparing the second and third column in table~2 we see that the $gg$ background
cross section is about one half of the total $q\bar q$ annihilation cross
section. 
\par 
In fig.~3,4 and 5 we present the transverse momentum distribution of the 
$b$--quark for $M_t = 175,\ 150$ and $200$ GeV, respectively,
at $\sqrt{s}=10$ and $14$ TeV.
In fig.~6 one can find the transverse momentum distribution of the $W^+$
for $M_t = 175$ GeV at both LHC energies.
The two double resonant contributions and the gluon--gluon fusion
background are shown separately and the full line gives the sum of the three
distributions.
Within the statistical errors of our Montecarlo the distributions of the two
double resonant channels differ only in overall normalization and have the same
shape. The $gg$ background $p^b_T$ spectrum is peaked at small transverse
momenta and shows a small hump at about $50$ GeV. This can be easily 
understood noticing that this channel is dominated
by diagrams, for instance diagram 6 and 7 in fig.~1, in which the two initial
state gluons split into $b\bar b$ pairs, producing $b$'s with small $p_T$.
A $b$ from one pair and a $\bar b$ from the other 
then further interact annihilating to a $W^+W^-$ pair
through the exchange of a $t$--channel top. The large mass of the top prevents
any substantial enhancement for small angle scattering and indeed the background
$p^W_T$ spectrum does not show any preference for small transverse momenta.
It is however softer than the corresponding double resonant distribution,
as can be seen if the different contributions are scaled to a common height,
say, of the maximum. The hump at large $p^b_T$ is due to the presence of
a single top in most of the background events.
\par
A more refined analysis of the influence of backgrounds on $t\bar t$ production
would require the introduction of a realistic set of experimental cuts. However,
as already mentioned, the vast majority of background events contain one
nearly on
shell $t$--quark and they would survive most selection procedures optimized
for the observation of top, leaving the signal to background ratio
practically unchanged. A possible catch in the preceding argument has to do with
the fact that the $b$--quark which is not produced in the decay of a top, has
a soft $p_T$ spectrum and some of the tracks from its own decay
might miss the vertex detector,
which is the most effective tool for $b$--tagging, whose geometrical
coverage is limited. This might reduce the $b$--tagging efficiency 
for this channel in comparison with the double resonant one where both $b$'s are
equally hard. It is however extremely difficult to
estimate this effect without a detailed simulation of $b$ hadronization and of
the full detector.
\par

\subsection*{Conclusions}
We have produced the complete matrix elements for $gg , q \bar q \rightarrow
b \bar b W^+W^-$. With them we have analyzed $t \bar t$ production at the
Tevatron and the LHC. With our approach we avoid
separating the production process from the subsequent decay. Finite width
effects, irreducible backgrounds and correlations between the two $t$-quark
decays are included in our treatment. 
\par
We found that contributions from irreducible backgrounds
provide about 1\% of the total $t\bar t$ cross section at the Tevatron 
and become more  relevant at the LHC, where they amount to about 5\% of
the total production rate. The reason for this difference is that single
resonant
channels are important for the $gg$ cross section and not for the 
$q \bar q$ processes. The the ${\cal O}(\alpha^4 )$ $q\bar q$ annihilation
cross section is about 2.5\% of the total at the Tevatron and is negligible at
the LHC.
\par

\vfill\eject

\vfill
\newpage
\begin{description}
\item[Table 1] Tree level total cross sections in $pb$
 for $p\bar p\ra b\bar bW^+W^-$ at
$\sqrt{s}=1.8$ TeV, $pp\ra b\bar bW^+W^-$ at
$\sqrt{s}=10$ TeV  and $pp\ra b\bar bW^+W^-$ at
$\sqrt{s}=14$ TeV  with $m_{top}=150,175,200$ GeV. The first and second
column refer to the contribution of double resonant diagrams and
to the contribution of single resonant diagrams and of the 
interference between single and double resonant diagrams in gluon-gluon fusion
respectively.
The third and fourth column report the contribution
from double resonant production via $q\bar q$ fusion into a gluon and
into $\gamma$ and $Z^0$ bosons respectively. 
The last column gives the sum of the four contribution.

\item[Table 2] Tree level total cross section in $pb$ and 
contribution in percentage of the most relevant channels to the total
cross section
for $pp\ra b\bar bW^+W^-$ at $\sqrt{s}=10$ TeV  and  at
$\sqrt{s}=14$ TeV  with $m_{top}=150,175,200$ GeV.
The first contribution arises from gluon fusion double resonant diagrams.
The second one is the contribution from single resonant diagrams and from the
interference between single and double resonant diagrams in the gluon fusion
channel, while the third one derives from quark
fusion double resonant diagrams. All other contributions are
below 1\% at tree level.

\end{description}

\vfill
\newpage

\bc
\begin{tabular}{|c|c|c|c|c|c|} \hline 

$m_{top}$ &$gg$&$gg$&$q\bar q$&$q\bar q$&total cross \\
 &$t\bar t$&bckgrd&$t\bar t$(strong)&$t\bar t$(el.weak)&section\\ \hline\hline
\multicolumn{6}{|c|}{$ \rule[-8 pt]{0 pt}{25 pt} p\bar p 
\hspace{.5 cm} \sqrt{s}=1.8$ TeV}  \\ \hline\hline
$150$ GeV&$1.219$&$0.075$&$6.53$&$0.157$&$7.98$ \\
$175$ GeV&$0.340$&$0.036$&$3.07$&$0.076$&$3.52$ \\
$200$ GeV&$0.101$&$0.018$&$1.49$&$0.037$&$1.64$ \\ \hline\hline
\multicolumn{6}{|c|}{$\rule[-8 pt]{0 pt}{25 pt} pp 
\hspace{.5 cm} \sqrt{s}=10$ TeV}  \\ \hline\hline
$150$ GeV&$418.$&$18.$&$63.2$&1.40&$500.$ \\
$175$ GeV&$201.$&$12.$&$35.3$&0.79&$249.$ \\
$200$ GeV&$104.$&$8.5$&$21.0$&0.48&$134.$ \\ \hline\hline
\multicolumn{6}{|c|}{$ \rule[-8 pt]{0 pt}{25 pt} pp 
\hspace{.5 cm}\sqrt{s}=14$ TeV}  \\ \hline\hline
$150$ GeV&$884.$&$38.$&$101.$&2.2&$1025.$ \\
$175$ GeV&$443.$&$26.$&$58.0$&1.3&$528.$ \\
$200$ GeV&$239.$&$19.$&$35.4$&0.8&$294.$ \\ \hline 
\end{tabular}\\
\vspace{.5 cm}
{\bf Tab.1}
\ec
\vfill
\bc
\begin{tabular}{|c|c|c|c|c|} \hline 
$m_{top}$&total cross&$gg\rightarrow b\bar b W^+W^-$&$gg\rightarrow
b\bar b W^+W^-$&$q \bar q\rightarrow b\bar b W^+W^-$ \\
 &section&signal&background&signal \\ \hline\hline
\multicolumn{5}{|c|}{$ \rule[-8 pt]{0 pt}{25 pt} pp 
\hspace{.5 cm} \sqrt{s}=10$ TeV}  \\ \hline\hline
$150$ GeV&$500.$ pb&$83\%$&$4\%$&$13\%$ \\
$175$ GeV&$249.$ pb&$81\%$&$5\%$&$14\%$ \\
$200$ GeV&$134.$ pb&$78\%$&$6\%$&$16\%$ \\ \hline\hline
\multicolumn{5}{|c|}{$ \rule[-8 pt]{0 pt}{25 pt}pp 
\hspace{.5 cm} \sqrt{s}=14$ TeV}  \\ \hline\hline
$150$ GeV&$1025.$ pb&$86\%$&$4\%$&$10\%$ \\
$175$ GeV&$528.$ pb&$84\%$&$5\%$&$11\%$ \\ 
$200$ GeV&$294.$ pb&$82\%$&$6\%$&$12\%$ \\ \hline 
\end{tabular} \\
\vspace{.5 cm}
{\bf Tab.2}
\ec

\vfill
\newpage
\subsection*{Figure Captions} 

\begin{description}

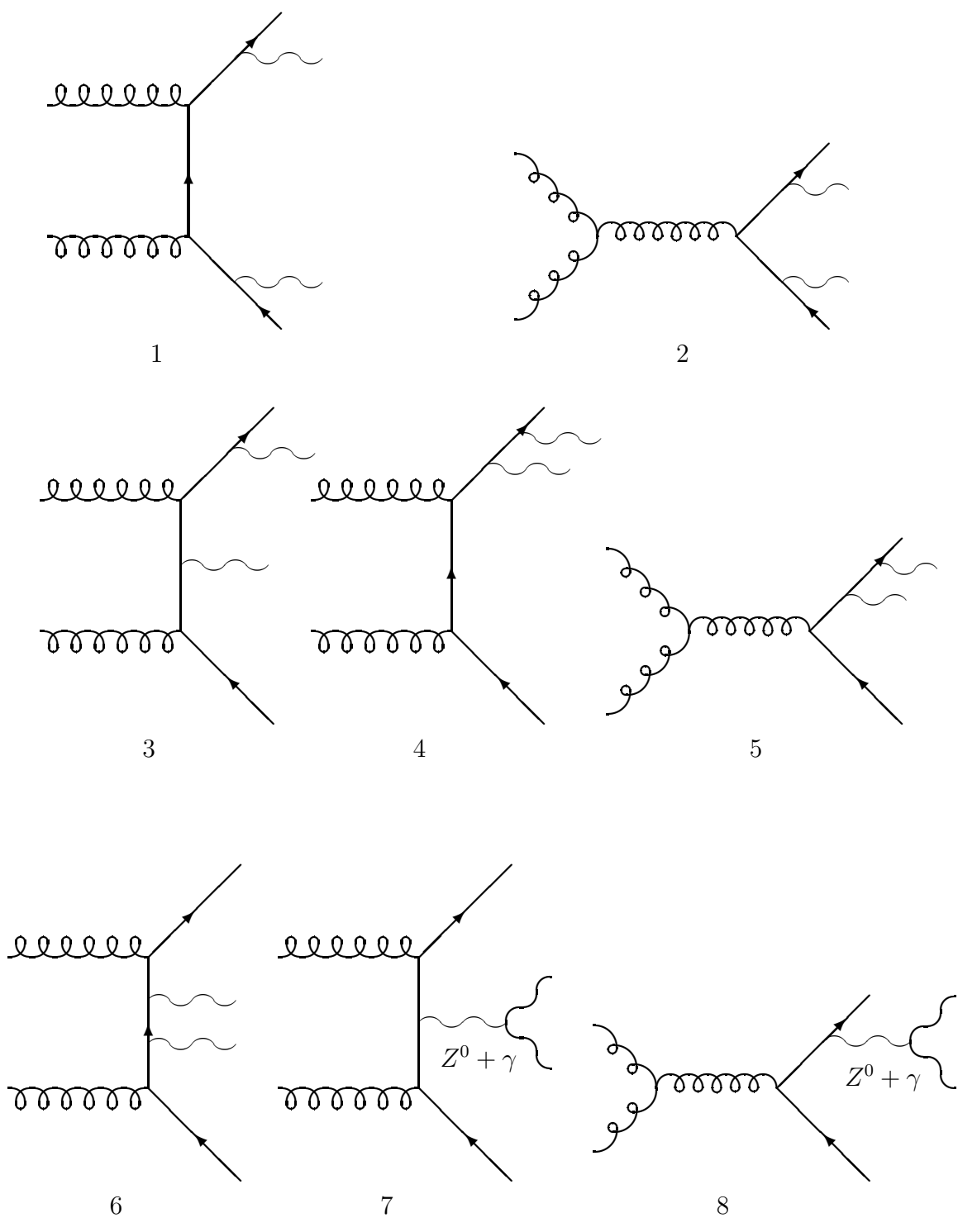
\item[Fig. 1]  Sample Feynman diagrams contributing in the lowest order to
$gg \ra b\bar bW^+W^-$. External wavy lines represent $W^\pm$ as
appropriate. Internal wavy lines represent $Z^0$ and $\gamma$.
Gluons permutation
are not shown. In diagrams (3)-(5)
only $t$-resonant diagrams are drawn; $\bar t$-resonant diagrams are obtained
replacing $b\leftrightarrow \bar b$ and $W^+ \leftrightarrow W^-$. 
All other non-resonant diagrams can be obtained from (7)
-(8) changing the
insertions of the internal vector boson lines ($Z_0$ and $\gamma$).
 SM Higgs contributions are not shown because their contribution is far below
 the accuracy reached in the present work.
 They can be obtained replacing internal neutral vector boson lines
 with scalar Higgs lines.

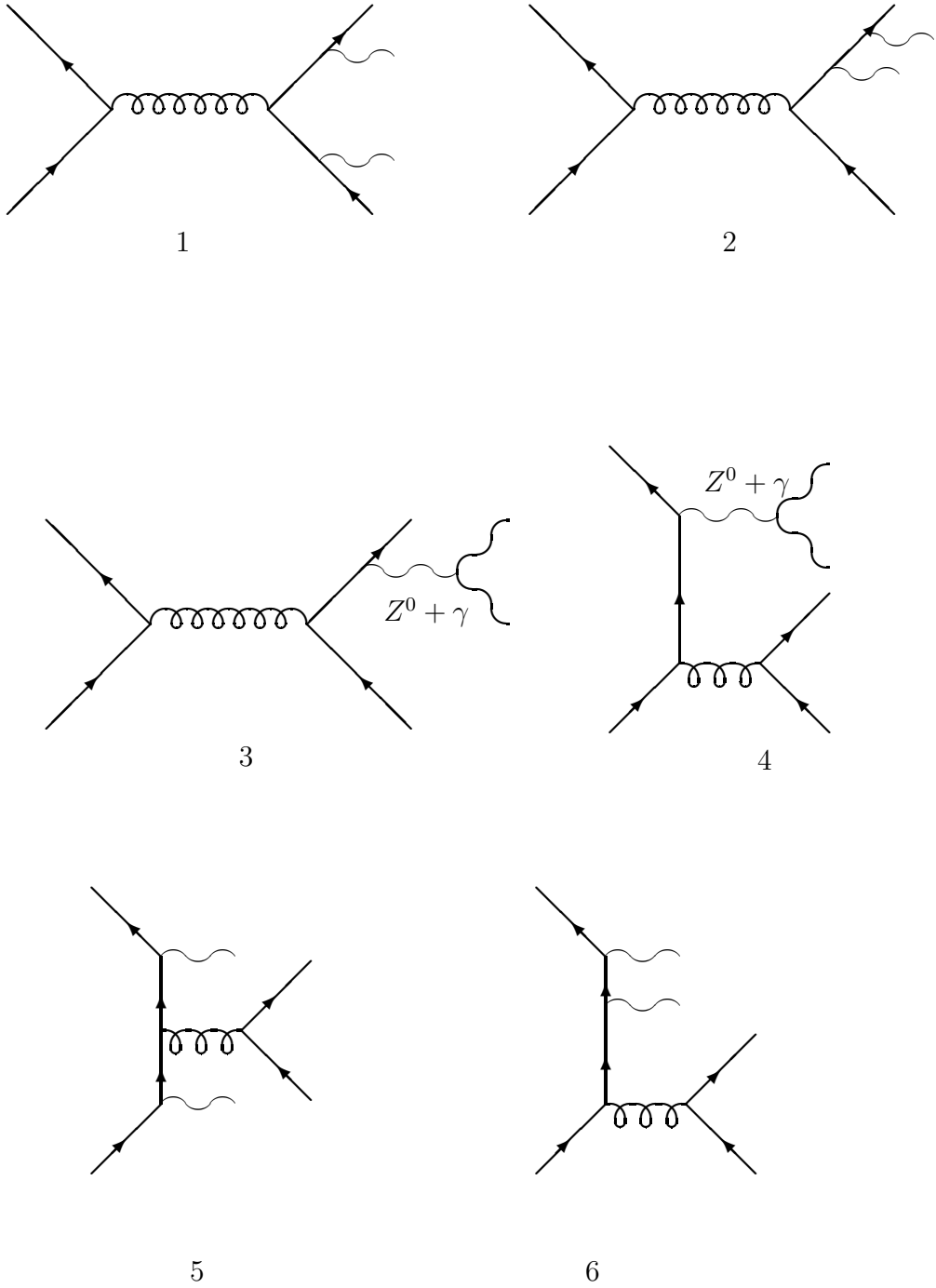
\item[Fig. 2]  Sample Feynman diagrams contributing in the lowest order to
$q\bar q \ra b\bar bW^+W^-$. External wavy lines represent $W^\pm$ as
appropriate. Internal wavy lines represent $Z^0$ and $\gamma$. 
Gluons permutation are not shown. 
$\bar t$-resonant diagrams are obtained replacing $b\leftrightarrow \bar b$
and $W^+ \leftrightarrow W^-$ in diagram (2).
All other non-resonant diagrams can be obtained from (3)
-(6) changing the insertions of the internal vector boson lines 
($Z_0$ and $\gamma$).
SM Higgs contributions are not shown because their contribution is far below
the accuracy reached in the present work.
They can be obtained replacing internal neutral vector boson lines
with scalar Higgs lines.

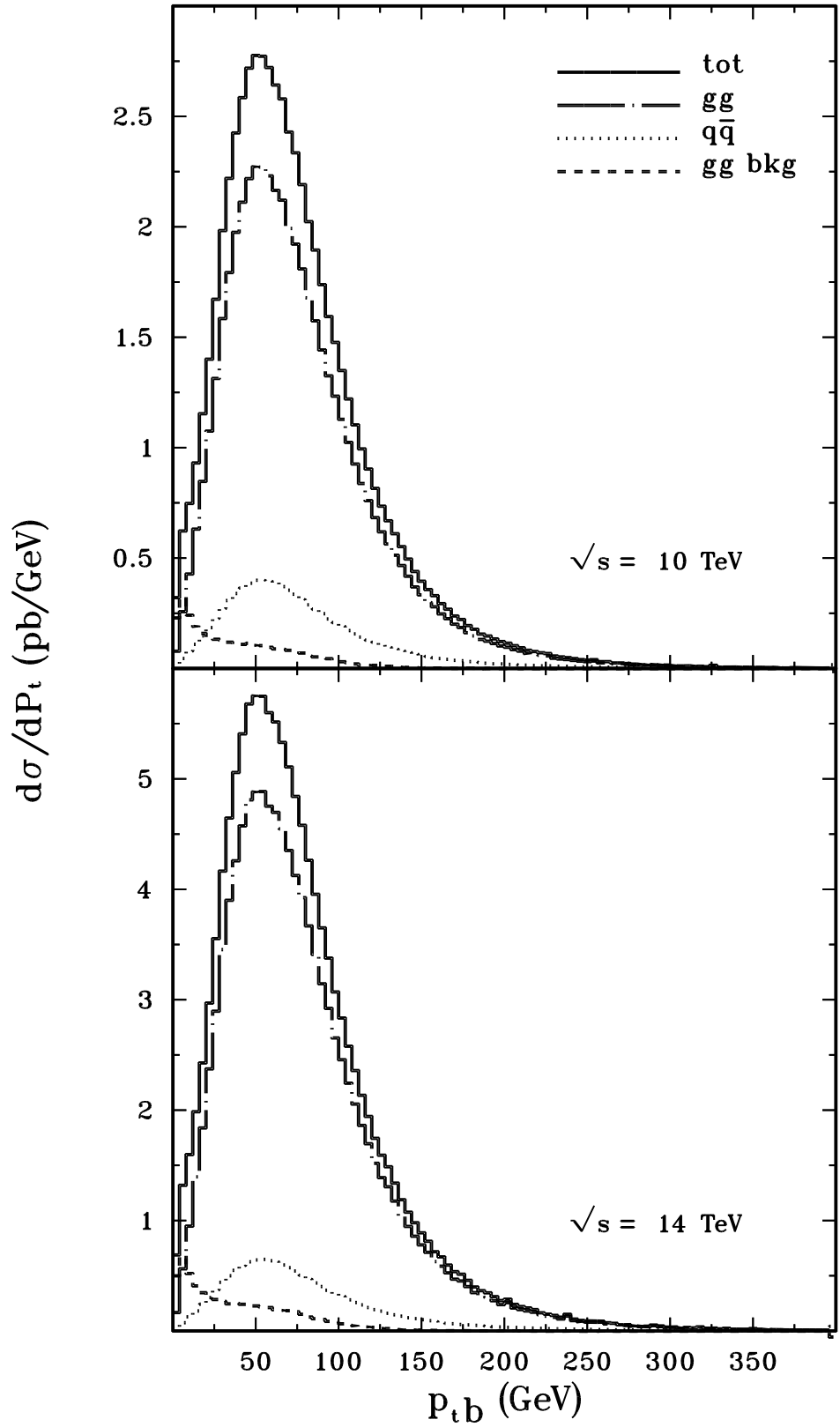
\item[Fig. 3] The distribution of transverse momentum of the
$b$ quark for $pp\ra b\bar
bW^+W^-$ with $m_{top}=175$ GeV at $\sqrt{s}=10$ TeV (Fig. 3a) and
$\sqrt{s}=14$ TeV (Fig. 3b). The dotted line represents the contribution from
double resonant diagrams in the quark fusion channel, while the dash-dotted line
is the contribution from double resonant diagrams in gluon fusion. The dashed
line is the "irreducible background" from single resonant diagrams and from the
interference between single and double resonant diagrams in the gluon fusion
channel. 
The continuous line is the sum of the three contributions.
 
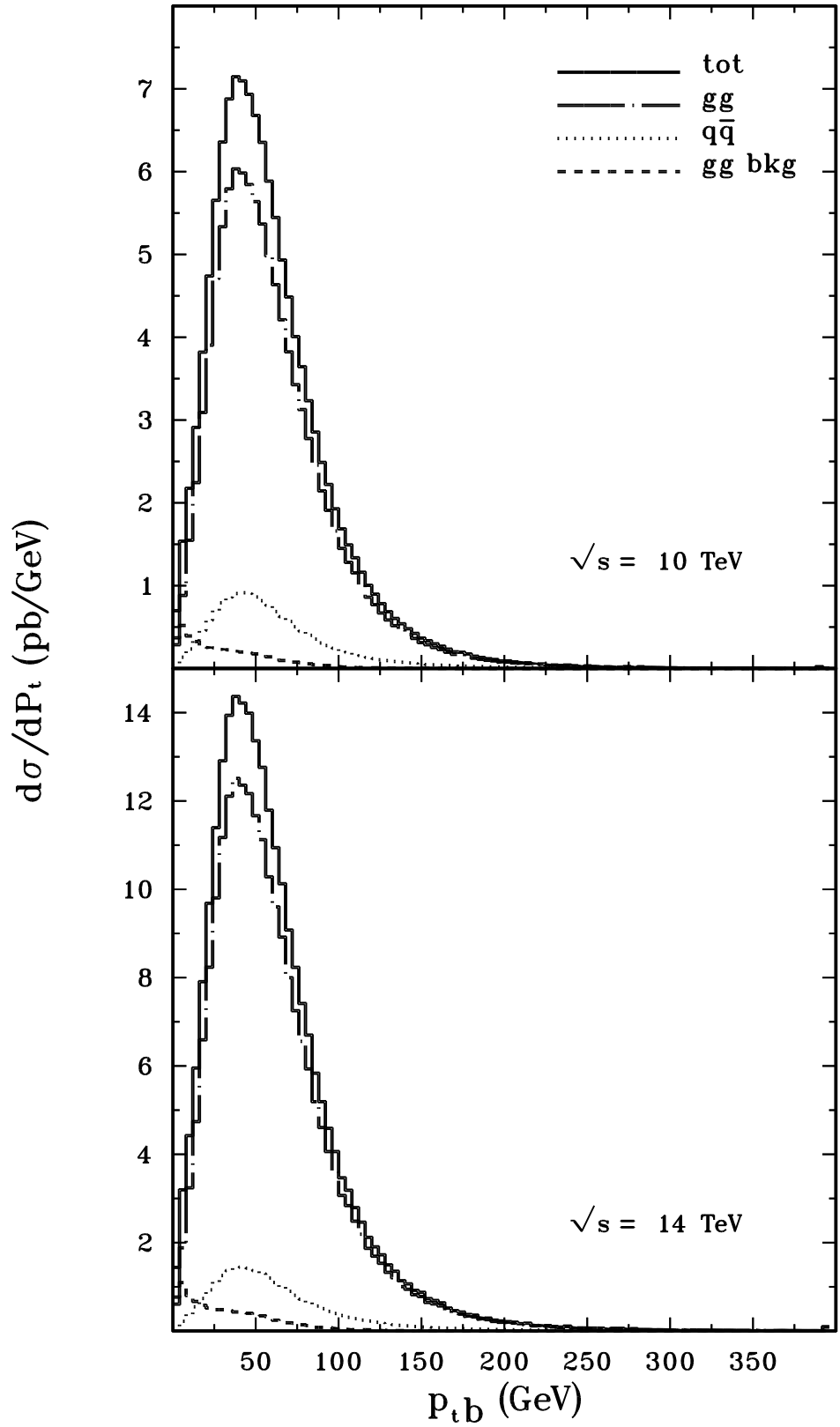
\item[Fig. 4] The distribution of transverse momentum of the
$b$ quark for $pp\ra b\bar
bW^+W^-$ with $m_{top}=150$ GeV at $\sqrt{s}=10$ TeV (Fig. 4a) and
$\sqrt{s}=14$ TeV (Fig. 4b). Symbols as in Fig.~3.

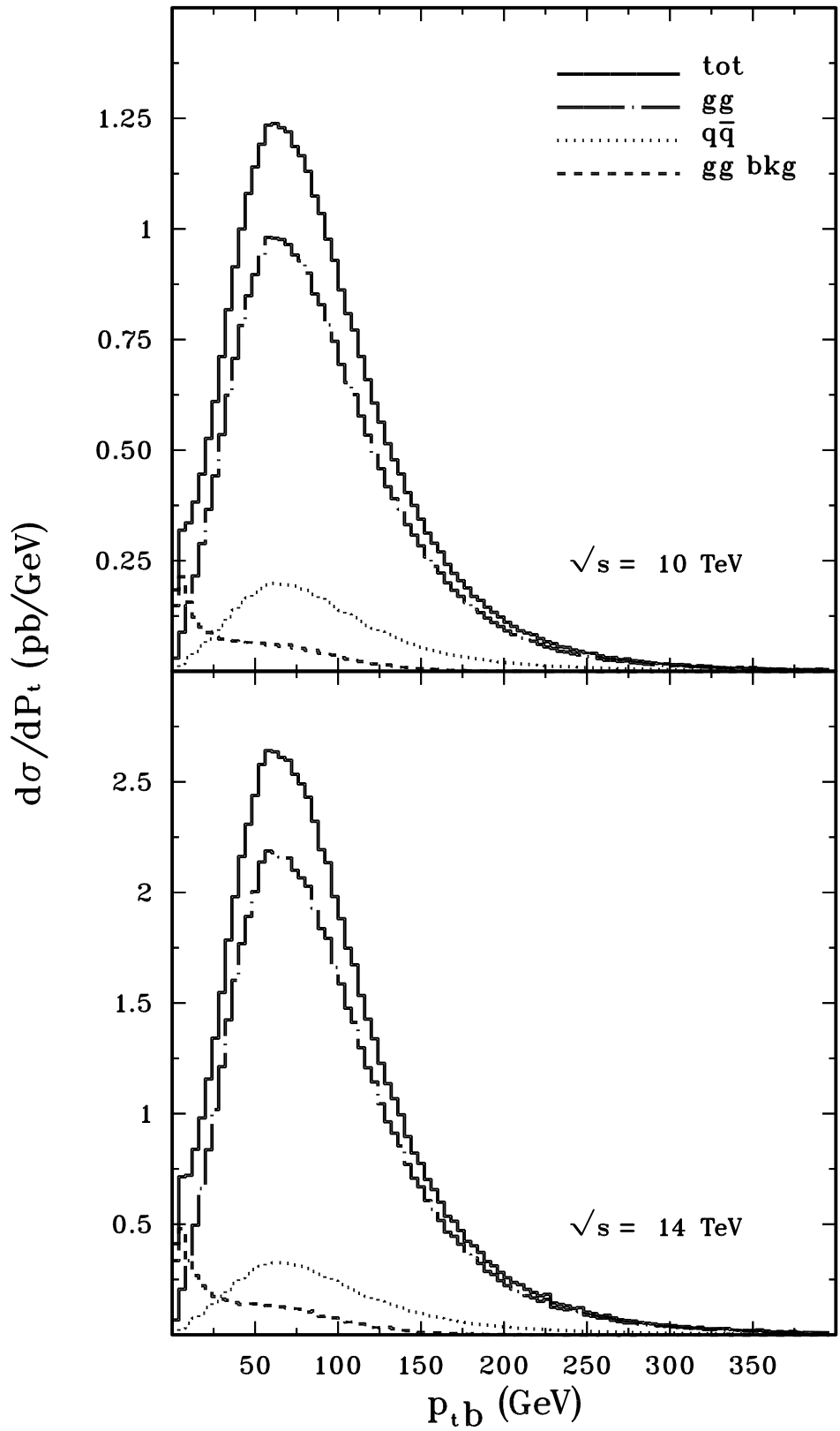
\item[Fig. 5] The distribution of transverse momentum of the
$b$ quark for $pp\ra b\bar
bW^+W^-$ with $m_{top}=200$ GeV at $\sqrt{s}=10$ TeV (Fig. 5a) and
$\sqrt{s}=14$ TeV (Fig. 5b). Symbols as in Fig.~3.

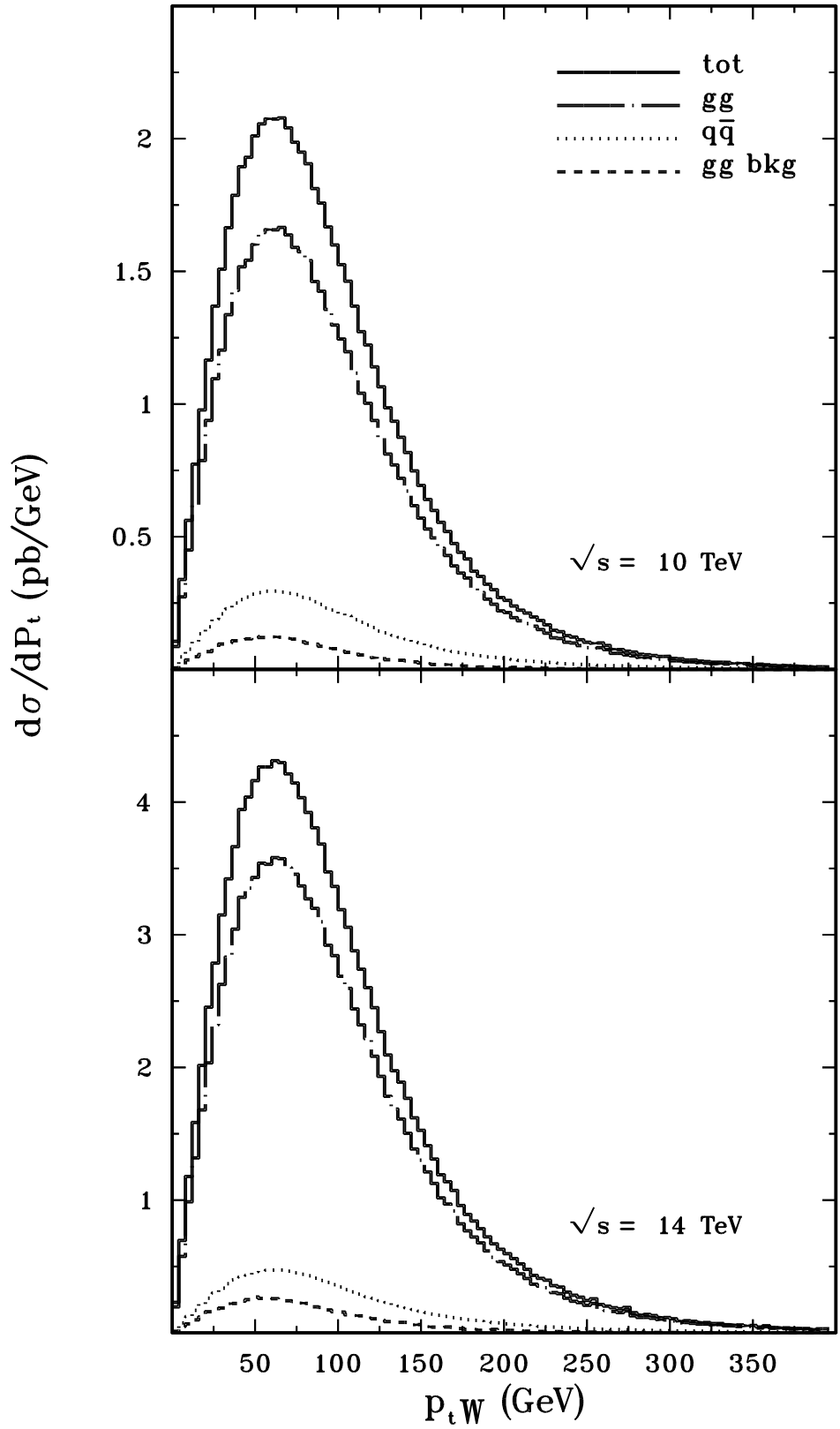
\item[Fig. 6] The distribution of transverse momentum of the $W^+$ 
for $pp\ra b\bar
bW^+W^-$ with $m_{top}=175$ GeV at $\sqrt{s}=10$ TeV (Fig. 6a)  and
$\sqrt{s}=14$ TeV (Fig. 6b). Symbols as in Fig.~3.

\end{description}

\newpage

\pagestyle{empty}
\thicklines
\bigphotons
\bc
\begin{picture}(15000,18000)
\drawline\fermion[\SW\REG](12000,18000)[3000]
\drawarrow[\NE\ATTIP](\particlemidx,\particlemidy)
\drawline\photon[\E\REG](\fermionbackx,\fermionbacky)[4]
\drawline\fermion[\SW\REG](\fermionbackx,\fermionbacky)[3000]
\drawline\gluon[\W\REG](\fermionbackx,\fermionbacky)[6]
\drawline\fermion[\S\REG](\fermionbackx,\fermionbacky)[6000]
\drawarrow[\N\ATTIP](\particlemidx,\particlemidy)
\drawline\gluon[\W\FLIPPED](\fermionbackx,\fermionbacky)[6]
\drawline\fermion[\SE\REG](\fermionbackx,\fermionbacky)[3000]
\drawline\photon[\E\REG](\fermionbackx,\fermionbacky)[4]
\drawline\fermion[\SE\REG](\fermionbackx,\fermionbacky)[3000]
\drawarrow[\NW\ATTIP](\particlemidx,\particlemidy)
\advance\fermionbacky by -1500
\put(6000,2000){1}
\end{picture}
\hskip 2 truecm
\vspace{-1. cm}
\begin{picture}(15000,12000)
\drawline\fermion[\SW\REG](16000,12000)[3000]
\drawarrow[\NE\ATTIP](\particlemidx,\particlemidy)
\drawline\photon[\E\REG](\fermionbackx,\fermionbacky)[3]
\drawline\fermion[\SW\REG](\fermionbackx,\fermionbacky)[3000]
\drawline\gluon[\W\FLIPPEDCENTRAL](\fermionbackx,\fermionbacky)[2]\gluonlink
\drawvertex\gluon[\W 3](\particlebackx,\particlebacky)[3]
\drawline\fermion[\SE\REG](\fermionbackx,\fermionbacky)[3000]
\drawline\photon[\E\REG](\fermionbackx,\fermionbacky)[3]
\drawline\fermion[\SE\REG](\fermionbackx,\fermionbacky)[3000]
\drawarrow[\NW\ATTIP](\particlemidx,\particlemidy)
\advance\fermionbacky by -1500
\put(9000,2000){2}
\end{picture} \\
\vspace{1. cm}
\begin{picture}(12000,18000)
\drawline\fermion[\SW\REG](12000,18000)[3000]
\drawarrow[\NE\ATTIP](\particlemidx,\particlemidy)
\drawline\photon[\E\REG](\fermionbackx,\fermionbacky)[4]
\drawline\fermion[\SW\REG](\fermionbackx,\fermionbacky)[3000]
\drawline\gluon[\W\REG](\fermionbackx,\fermionbacky)[6]
\drawline\fermion[\S\REG](\fermionbackx,\fermionbacky)[3000]
\drawline\photon[\E\REG](\fermionbackx,\fermionbacky)[4]
\drawline\fermion[\S\REG](\fermionbackx,\fermionbacky)[3000]
\drawline\gluon[\W\FLIPPED](\fermionbackx,\fermionbacky)[6]
\drawline\fermion[\SE\REG](\fermionbackx,\fermionbacky)[6000]
\drawarrow[\NW\ATTIP](\particlemidx,\particlemidy)
\advance\fermionbacky by -1500
\put(6000,2000){3}
\end{picture}
\begin{picture}(12000,18000)
\drawline\fermion[\SW\REG](12000,18000)[2000]
\drawarrow[\NE\ATTIP](\particlemidx,\particlemidy)
\drawline\photon[\E\REG](\fermionbackx,\fermionbacky)[4]
\drawline\fermion[\SW\REG](\fermionbackx,\fermionbacky)[2000]
\drawline\photon[\E\REG](\fermionbackx,\fermionbacky)[4]
\drawline\fermion[\SW\REG](\fermionbackx,\fermionbacky)[2000]
\drawline\gluon[\W\REG](\fermionbackx,\fermionbacky)[6]
\drawline\fermion[\S\REG](\fermionbackx,\fermionbacky)[6000]
\drawarrow[\N\ATTIP](\particlemidx,\particlemidy)
\drawline\gluon[\W\FLIPPED](\fermionbackx,\fermionbacky)[6]
\drawline\fermion[\SE\REG](\fermionbackx,\fermionbacky)[6000]
\drawarrow[\NW\ATTIP](\particlemidx,\particlemidy)
\advance\fermionbacky by -1500
\put(6000,2000){4}
\end{picture}
\begin{picture}(12000,12000)
\drawline\fermion[\SW\REG](16000,12000)[2000]
\drawarrow[\NE\ATTIP](\particlemidx,\particlemidy)
\drawline\photon[\E\REG](\fermionbackx,\fermionbacky)[3]
\drawline\fermion[\SW\REG](\fermionbackx,\fermionbacky)[2000]
\drawline\photon[\E\REG](\fermionbackx,\fermionbacky)[3]
\drawline\fermion[\SW\REG](\fermionbackx,\fermionbacky)[2000]
\drawline\gluon[\W\FLIPPEDCENTRAL](\fermionbackx,\fermionbacky)[1]\gluonlink
\drawvertex\gluon[\W 3](\particlebackx,\particlebacky)[3]
\drawline\fermion[\SE\REG](\fermionbackx,\fermionbacky)[6000]
\drawarrow[\NW\ATTIP](\particlemidx,\particlemidy)
\advance\fermionbacky by -1500
\put(9000,2000){5}
\end{picture} \\
\vspace{1. cm}
\begin{picture}(12000,18000)
\drawline\fermion[\SW\REG](12000,18000)[6000]
\drawarrow[\NE\ATTIP](\particlemidx,\particlemidy)
\drawline\gluon[\W\REG](\fermionbackx,\fermionbacky)[6]
\drawline\fermion[\S\REG](\fermionbackx,\fermionbacky)[2000]
\drawline\photon[\E\REG](\fermionbackx,\fermionbacky)[4]
\drawline\fermion[\S\REG](\fermionbackx,\fermionbacky)[2000]
\drawarrow[\N\ATTIP](\particlemidx,\particlemidy)
\drawline\photon[\E\REG](\fermionbackx,\fermionbacky)[4]
\drawline\fermion[\S\REG](\fermionbackx,\fermionbacky)[2000]
\drawline\gluon[\W\FLIPPED](\fermionbackx,\fermionbacky)[6]
\drawline\fermion[\SE\REG](\fermionbackx,\fermionbacky)[6000]
\drawarrow[\NW\ATTIP](\particlemidx,\particlemidy)
\advance\fermionbacky by -1500
\put(6000,2000){6}
\end{picture}
\begin{picture}(12000,18000)
\drawline\fermion[\SW\REG](12000,18000)[6000]
\drawarrow[\NE\ATTIP](\particlemidx,\particlemidy)
\drawline\gluon[\W\REG](\fermionbackx,\fermionbacky)[6]
\drawline\fermion[\S\REG](\fermionbackx,\fermionbacky)[3000]
\drawline\photon[\E\REG](\fermionbackx,\fermionbacky)[1]
\drawvertex\photon[\E 3](\particlebackx,\particlebacky)[3]
\advance\vertexoney by -2000
\put(\vertexonex,\vertexoney){$Z^0+\gamma$}
\drawline\fermion[\S\REG](\fermionbackx,\fermionbacky)[3000]
\drawline\gluon[\W\FLIPPED](\fermionbackx,\fermionbacky)[6]
\drawline\fermion[\SE\REG](\fermionbackx,\fermionbacky)[6000]
\drawarrow[\NW\ATTIP](\particlemidx,\particlemidy)
\advance\fermionbacky by -1500
\put(6000,2000){7}
\end{picture}
\begin{picture}(15000,12000)
\drawline\fermion[\SW\REG](16000,12000)[3000]
\drawarrow[\NE\ATTIP](\particlemidx,\particlemidy)
\drawline\photon[\E\REG](\fermionbackx,\fermionbacky)[1]
\drawvertex\photon[\E 3](\particlebackx,\particlebacky)[3]
\advance\vertexoney by -2000
\put(\vertexonex,\vertexoney){$Z^0+\gamma$}
\drawline\fermion[\SW\REG](\fermionbackx,\fermionbacky)[3000]
\drawline\gluon[\W\FLIPPEDCENTRAL](\fermionbackx,\fermionbacky)[2]\gluonlink
\drawvertex\gluon[\W 3](\particlebackx,\particlebacky)[2]
\drawline\fermion[\SE\REG](\fermionbackx,\fermionbacky)[6000]
\drawarrow[\NW\ATTIP](\particlemidx,\particlemidy)
\advance\fermionbacky by -1500
\put(9000,2000){8}
\end{picture} \\
\ec
\vspace{1. cm}
\bc
{\bf Fig.1}
\ec
\eject

\pagestyle{empty}

\thicklines
\bigphotons
\bc
\begin{picture}(18000,12000)
\drawline\fermion[\SW\REG](16000,12000)[3000]
\drawarrow[\NE\ATTIP](\particlemidx,\particlemidy)
\drawline\photon[\E\REG](\fermionbackx,\fermionbacky)[3]
\drawline\fermion[\SW\REG](\fermionbackx,\fermionbacky)[3000]
\drawline\gluon[\W\FLIPPEDCENTRAL](\fermionbackx,\fermionbacky)[6]
\drawline\fermion[\NW\REG](\gluonbackx,\gluonbacky)[6000]
\drawarrow[\NW\ATTIP](\particlemidx,\particlemidy)
\drawline\fermion[\SW\REG](\gluonbackx,\gluonbacky)[6000]
\drawarrow[\NE\ATTIP](\particlemidx,\particlemidy)
\drawline\fermion[\SE\REG](\gluonfrontx,\gluonfronty)[3000]
\drawline\photon[\E\REG](\fermionbackx,\fermionbacky)[3]
\drawline\fermion[\SE\REG](\fermionbackx,\fermionbacky)[3000]
\drawarrow[\NW\ATTIP](\particlemidx,\particlemidy)
\advance\fermionbacky by -1500
\put(8000,2000){1}
\end{picture}
\hskip 1. truecm
\begin{picture}(18000,12000)
\drawline\fermion[\SW\REG](16000,12000)[2000]
\drawarrow[\NE\ATTIP](\particlemidx,\particlemidy)
\drawline\photon[\E\REG](\fermionbackx,\fermionbacky)[3]
\drawline\fermion[\SW\REG](\fermionbackx,\fermionbacky)[2000]
\drawline\photon[\E\REG](\fermionbackx,\fermionbacky)[3]
\drawline\fermion[\SW\REG](\fermionbackx,\fermionbacky)[2000]
\drawline\gluon[\W\FLIPPEDCENTRAL](\fermionbackx,\fermionbacky)[6]
\drawline\fermion[\NW\REG](\gluonbackx,\gluonbacky)[6000]
\drawarrow[\NW\ATTIP](\particlemidx,\particlemidy)
\drawline\fermion[\SW\REG](\gluonbackx,\gluonbacky)[6000]
\drawarrow[\NE\ATTIP](\particlemidx,\particlemidy)
\drawline\fermion[\SE\REG](\gluonfrontx,\gluonfronty)[6000]
\drawarrow[\NW\ATTIP](\particlemidx,\particlemidy)
\advance\fermionbacky by -1500
\put(9000,2000){2}
\end{picture}\\
\vspace{1. cm}
\begin{picture}(18000,12000)
\drawline\fermion[\SW\REG](16000,12000)[3000]
\drawarrow[\NE\ATTIP](\particlemidx,\particlemidy)
\drawline\photon[\E\REG](\fermionbackx,\fermionbacky)[1]
\drawvertex\photon[\E 3](\particlebackx,\particlebacky)[3]
\advance\vertexoney by -2000
\put(\vertexonex,\vertexoney){$Z^0+\gamma$}
\drawline\fermion[\SW\REG](\fermionbackx,\fermionbacky)[3000]
\drawline\gluon[\W\FLIPPEDCENTRAL](\fermionbackx,\fermionbacky)[6]
\drawline\fermion[\NW\REG](\gluonbackx,\gluonbacky)[6000]
\drawarrow[\NW\ATTIP](\particlemidx,\particlemidy)
\drawline\fermion[\SW\REG](\gluonbackx,\gluonbacky)[6000]
\drawarrow[\NE\ATTIP](\particlemidx,\particlemidy)
\drawline\fermion[\SE\REG](\gluonfrontx,\gluonfronty)[6000]
\drawarrow[\NW\ATTIP](\particlemidx,\particlemidy)
\advance\fermionbacky by -1500
\put(9000,2000){3}
\end{picture}
\hskip 2. truecm
\begin{picture}(12000,18000)
\drawline\fermion[\SE\REG](0,15000)[4000]
\drawarrow[\NW\ATTIP](\particlemidx,\particlemidy)
\drawline\photon[\E\REG](\fermionbackx,\fermionbacky)[1]
\drawvertex\photon[\E 3](\particlebackx,\particlebacky)[3]
\advance\vertexoney by 1000
\put(\vertexonex,\vertexoney){$Z^0+\gamma$}
\drawline\fermion[\S\REG](\fermionbackx,\fermionbacky)[6000]
\drawarrow[\N\ATTIP](\particlemidx,\particlemidy)
\drawline\gluon[\E\REG](\fermionbackx,\fermionbacky)[3]
\drawline\fermion[\NE\REG](\gluonbackx,\gluonbacky)[4000]
\drawarrow[\NE\ATTIP](\particlemidx,\particlemidy)
\drawline\fermion[\SE\REG](\gluonbackx,\gluonbacky)[4000]
\drawarrow[\NW\ATTIP](\particlemidx,\particlemidy)
\drawline\fermion[\SW\REG](\gluonfrontx,\gluonfronty)[4000]
\drawarrow[\NE\ATTIP](\particlemidx,\particlemidy)
\advance\fermionbacky by -1500
\put(6000,\fermionbacky){4}
\end{picture}\\
\vspace{1. cm}
\begin{picture}(12000,18000)
\drawline\fermion[\SE\REG](0,18000)[4000]
\drawarrow[\NW\ATTIP](\particlemidx,\particlemidy)
\drawline\photon[\E\REG](\fermionbackx,\fermionbacky)[3]
\drawline\fermion[\S\REG](\fermionbackx,\fermionbacky)[3000]
\drawarrow[\N\ATTIP](\particlemidx,\particlemidy)
\drawline\gluon[\E\REG](\fermionbackx,\fermionbacky)[3]
\drawline\fermion[\NE\REG](\gluonbackx,\gluonbacky)[4000]
\drawarrow[\NE\ATTIP](\particlemidx,\particlemidy)
\drawline\fermion[\SE\REG](\gluonbackx,\gluonbacky)[4000]
\drawarrow[\NW\ATTIP](\particlemidx,\particlemidy)
\drawline\fermion[\S\REG](\gluonfrontx,\gluonfronty)[3000]
\drawarrow[\N\ATTIP](\particlemidx,\particlemidy)
\drawline\photon[\E\REG](\fermionbackx,\fermionbacky)[3]
\drawline\fermion[\SW\REG](\fermionbackx,\fermionbacky)[4000]
\drawarrow[\NE\ATTIP](\particlemidx,\particlemidy)
\advance\fermionbacky by -1500
\put(4000,2000){5}
\end{picture}
\hskip 2. truecm
\begin{picture}(12000,18000)
\drawline\fermion[\SE\REG](0,18000)[4000]
\drawarrow[\NW\ATTIP](\particlemidx,\particlemidy)
\drawline\photon[\E\REG](\fermionbackx,\fermionbacky)[3]
\drawline\fermion[\S\REG](\fermionbackx,\fermionbacky)[2000]
\drawarrow[\N\ATTIP](\particlemidx,\particlemidy)
\drawline\photon[\E\REG](\fermionbackx,\fermionbacky)[3]
\drawline\fermion[\S\REG](\fermionbackx,\fermionbacky)[4000]
\drawarrow[\N\ATTIP](\particlemidx,\particlemidy)
\drawline\gluon[\E\REG](\fermionbackx,\fermionbacky)[3]
\drawline\fermion[\NE\REG](\gluonbackx,\gluonbacky)[4000]
\drawarrow[\NE\ATTIP](\particlemidx,\particlemidy)
\drawline\fermion[\SE\REG](\gluonbackx,\gluonbacky)[4000]
\drawarrow[\NW\ATTIP](\particlemidx,\particlemidy)
\drawline\fermion[\SW\REG](\gluonfrontx,\gluonfronty)[4000]
\drawarrow[\NE\ATTIP](\particlemidx,\particlemidy)
\advance\fermionbacky by -1500
\put(2000,2000){6}
\end{picture}
\ec
\vspace{2. cm}
\bc
{\bf Fig.2}
\ec
\eject

\end{document}